\documentclass[12pt,aps,prd,showpacs,amsmath,amssymb]{revtex4}
\input epsf
\begin{document}

\title{Semileptonic $B$ Meson Decays Into A Highly Excited Charmed Meson Doublet}

\author{Long-Fei Gan\footnote{lfgan@nudt.edu.cn}}
\author{Ming-Qiu Huang}

\affiliation{Department of Physics, National University of Defense
Technology, Hunan 410073, China}

\date{\today}

\begin{abstract}
We study the heavy quark effective theory prediction for
semileptonic $B$ decays into an orbital excited $F$-wave charmed
doublet, the ($2^{+}$, $3^{+}$) states ($D^{*'}_{2}$, $D_{3}$), at
the leading order of heavy quark expansion. The corresponding
universal form factor is estimated by using the QCD sum rule method.
The decay rates we predict are $\Gamma_{B\rightarrow
D^{*'}_{2}\ell\overline{\nu}}=1.85\times10^{-19} \mbox{GeV}$ and
$\Gamma_{B\rightarrow D_{3}\ell\overline{\nu}}=1.78\times10^{-19}
\mbox{GeV}$. The branching ratios are $\mathcal {B}(B\rightarrow
D_{2}^{*'}\ell\overline{\nu})=4.6\times10^{-7}$ and $\mathcal
{B}(B\rightarrow D_{3}\ell\overline{\nu})=4.4\times10^{-7}$,
respectively.
\end{abstract}

\maketitle

The semileptonic $B$ decay processes with charmed mesons in their
final states have attracted attention in recent years for the
important role that they played in the determination of the
Cabibbo-Kobayashi-Maskawa (CKM) matrix elements, such as $|V_{cb}|$.
They also provide a useful testing ground for the strong interaction
as well as heavy quark symmetry. Theoretically, the heavy quark
effective theory (HQET) \cite{Neu1} has been well developed and
widely applied to these processes. In HQET, the heavy quark
spin-flavor symmetry brings much convenience to understanding the
heavy-light mesons and their decays. On the one hand, the
heavy-light meson spectrum can be organized according to the parity
$P$ and the total angular momentum $s_{l}$ of the light degrees of
freedom. Coupling the spin of the light degrees of freedom $s_{l}$
with the spin of a heavy quark $s_{Q}=1/2$ yields a doublet of meson
states with a total spin $s=s_{l}\pm1/2$. The spectra of the
heavy-light mesons were given early in Ref. \cite{God} by a
relativistic quark model. The properties of the ground and low-lying
states have been extensively studied using different approaches
during the past few years \cite{Dai,Dai1,Xia,Wei}. On the other
hand, the weak transition matrix elements describing semileptonic
$B$ decays into a charmed meson can be parametrized, respectively,
by one universal Isgur-Wise function at the leading order of heavy
quark expansion \cite{Isg,Wis}. This simplifies the theoretical
calculation of these processes dramatically. Experimentally, some of
the $S$-wave and $P$-wave charmed states have been observed so far.
Their masses and quantum numbers have been well confirmed. In
addition, besides the measurement of the $B\rightarrow
D^{(*)}\ell\overline{\nu}$ decay which accounts for most of the
inclusive semileptonic $B$ decay branching ratio, more and more
subdominant decay channels have been measured with an increasing
accuracy\cite{Cle,Ale,Aub1}.

Semileptonic $B$ decays into low-lying charmed mesons were studied
in the past \cite{Wis,Neu,Hua1,Ebe1}. In our previous work, we have
given the HQET predictions for the semileptonic $B$ decays into the
$D$-wave charmed meson doublets \cite{Gan}. We find that the
branching ratios are of the same order as those of the CKM
suppressed $b\rightarrow u\ell\nu$ processes. So we can expect that,
with the increasing of the orbital angular momenta of final excited
charmed mesons, the dynamical suppression becomes severer than the
CKM suppression. To make this clear, we study the HQET predictions
for the semileptonic $B$ decays into the next heavy meson doublet,
the $s_{l}^{P}=\frac{5}{2}^{+}$ doublet which comprises two mesons,
($D^{*'}_{2}$, $D_{3}$), with $J^{P}=2^{+}$ and $J^{P}=3^{+}$.

The theoretical description of semileptonic decays involves the
matrix elements of vector and axial vector currents ( $V^{\mu} =
\overline{c}\gamma^{\mu}b$ and $A^{\mu} =
\overline{c}\gamma^{\mu}\gamma_{5}b$ ) between $B$ mesons and $D$
mesons, which can be parametrized into one universal form factor at
the leading order of the heavy quark expansion by applying the trace
formalism \cite{Fal}. For the processes $B\rightarrow (D_{2}^{*'},
D_{3})\ell\overline{\nu}$, these matrix elements turn out to be
\begin{eqnarray}\label{matrix1}
\langle D_{2}^{*'}(v^{'},\varepsilon)|(V-A)^{\mu}|B(v)\rangle&=&
\sqrt{\frac{5}{3}}\sqrt{m_{B}m_{D_{2}}}\tau(y)
\varepsilon^{*}_{\alpha\beta}v^{\alpha}[\frac{2(y^{2}-1)}{5}g^{\mu\beta}+v^{\beta}v^{\mu}
-\frac{2y+3}{5}v^{\beta}v^{'\mu}\nonumber\\&+&i\frac{2(y-1)}{5}\epsilon^{\mu\lambda\beta\rho}v_{\lambda}v^{'}_{\rho}],
\end{eqnarray}
\begin{equation}\label{matrix2}
\langle D_{3}(v^{'},\varepsilon)|(V-A)^{\mu}|B(v)\rangle =
\sqrt{m_{B}m_{D_{3}^{*}}} \tau(y)
\varepsilon^{*}_{\alpha\beta\lambda}v^{\alpha}v^{\beta}[g^{\mu\lambda}(y-1)-v^{\lambda}v^{'\mu}
-i\epsilon^{\mu\lambda\rho\tau}v_{\rho}v^{'}_{\tau}],
\end{equation}
where $y=v\cdot v^{'}$, and $\varepsilon^{*}_{\alpha\beta}$,
$\varepsilon^{*}_{\alpha\beta\lambda}$ are the polarization tensors
of these mesons. $\tau(y)$ is the Isgur-Wise function which can be
estimated by using the QCD sum rule method \cite{Shi}.

For this purpose, some appropriate interpolating currents are needed
to represent these states. Here we adopt the interpolating currents
proposed in Ref. \cite{Dai} based on the study of the Bethe-Salpeter
equation for heavy mesons. For the $F$-wave meson doublet with
$s_{l}^{P}=\frac{5}{2}^{+}$, the currents are given as follows:
\begin{equation}\label{current5}
J^{\dag\alpha\beta}_{2,+,5/2}=i\sqrt{\frac{5}{6}}T^{\alpha\beta,\mu\nu}\overline{h}_{v}
(D_{t\nu}D_{t\mu}-\frac{2}{5}D_{t\nu}\gamma_{t\mu}\rlap/D_{t})\rlap/D_{t}q,
\end{equation}
\begin{equation}\label{current6}
J^{\dag\alpha\beta\lambda}_{3,+,5/2}=\frac{i}{\sqrt{2}}T^{\alpha\beta\lambda,\mu\nu\sigma}\overline{h}_{v}\gamma_{5}\gamma_{t\mu}
D_{t\nu}D_{t\sigma}\rlap/D_{t}q,
\end{equation}
where $h_{v}$ is the generic velocity-dependent heavy quark
effective field in HQET and $q$ denotes the light quark field;
$D_{t\mu}=D_{\mu}-v_{\mu}(v\cdot D)$ is the transverse component of
the covariant derivative with respect to the velocity of the meson,
where $\gamma_{t\mu}=\gamma_{\mu}-\rlap/vv_{\mu}$ is the transverse
component of $\gamma_{\mu}$ with respect to the heavy quark
velocity. The tensors $T^{\alpha\beta,\mu\nu}$ and
$T^{\alpha\beta\lambda,\mu\nu\sigma}$ are used to symmetrize indices
\cite{Dai,Gan}. These currents have nice properties: they have
nonvanishing projection only to the corresponding states of the HQET
in the $m_{Q}\rightarrow\infty$ limit, without mixing with states of
the same quantum number but different $s_{l}$. Thus we can define
one-particle-current couplings as follows:
\begin{equation}\label{const1}
\langle D^{*'}_{2}(v,\varepsilon)|J^{\alpha\beta}|0\rangle=
f_{2}\sqrt{m_{D^{*'}_{2}}}\varepsilon^{*\alpha\beta}, \text{ for}
J^{P}=2^{+};
\end{equation}
\begin{equation}\label{const2}
\langle D_{3}(v,\varepsilon)|J^{\alpha\beta\lambda}|0\rangle=
f_{3}\sqrt{m_{D_{3}}}\varepsilon^{*\alpha\beta\lambda}, \text{ for}
J^{P}=3^{+}.
\end{equation}
The decay constants $f_{2}$ and $f_{3}$ are low-energy parameters
which are determined by the dynamics of the light degrees of
freedom. Hereafter, we denote them as $f_{+,\frac{5}{2}}$ in common.

The decay constants of heavy-light mesons can be estimated from
two-point sum rules. Here we list them after the Borel
transformation as follows. For the initial ground-state $B$ meson we
consider, the sum rule for the correlator of two heavy-light
currents of the ground-state is \cite{Hua1}
\begin{equation}\label{consrule1}
f^{2}_{-,\frac{1}{2}}e^{-2\bar{\Lambda}_{-,1/2}/T}=\frac{3}{16\pi^{2}}\int_{0}^{\omega_{c0}}
\omega^{2}e^{-\omega/T}d\omega-\frac{1}{2}\langle\bar{q}q\rangle(1-\frac{m^{2}_{0}}{4T^{2}}).
\end{equation}
For the final $s_{l}^{P}=\frac{5}{2}^{+}$ doublet, the corresponding
two-point sum rule is
\begin{eqnarray}\label{consrule2}
f^{2}_{+,\frac{5}{2}}e^{-2\bar{\Lambda}_{+,5/2}/T}
&=&\frac{1}{5\times2^{9}\pi^{2}}\int_{0}^{\omega_{c1}}
\omega^{8}e^{-\omega/T}d\omega-\frac{133}{9\times2^{10}}\int_{0}^{\omega_{c1}}
\omega^{4}e^{-\omega/T}d\omega\langle\frac{\alpha_{s}}{\pi}GG\rangle \nonumber \\
&=& \mathcal{F}(\omega_{c1},T).
\end{eqnarray}
The binding energy $\bar{\Lambda}$ of the F-wave meson states can be
estimated from Eq.(\ref{consrule2}) by taking the derivative about
$(-\frac{1}{T})$ on both sides, that is,
\begin{equation}\label{massrule}
\bar{\Lambda}_{+,5/2}=\frac{1}{2\mathcal{F}(\omega_{c1},T)}\frac{d\mathcal{F}(\omega_{c1},T)}{d(-\frac{1}{T})}.
\end{equation}

One can also estimate the universal form factor $\tau(y)$ via
three-point function sum rules by considering the three-point
correlator
\begin{equation}\label{rule1}
i^{2}\int d^{4}xd^{4}ze^{i(k^{'}\cdot x-k\cdot
z)}\langle0|T[J^{\alpha\beta}_{2,+}(x)
J^{\mu(v,v^{'})}_{V,A}(0)J^{\dag}_{0,-}(z)|0\rangle=\Gamma(\omega,\omega^{'},y)\mathcal
{L}^{\mu\alpha\beta}_{V,A},
\end{equation}
where $J^{\mu(v,v^{'})}_{V}=h(v^{'})\gamma^{\mu}h(v)$ and
$J^{\mu(v,v^{'})}_{A}=h(v^{'})\gamma^{\mu}\gamma_{5}h(v)$. The
variables $k$($=P-m_{b}v$) and $k^{'}$($=P'-m_{c}v'$) denote
residual ``off-shell" momenta of the initial and final meson states,
respectively. For heavy quarks in bound states they are typically of
order $\Lambda_{QCD}$ and remain finite in the heavy quark limit.
$\Gamma(\omega,\omega^{'},y)$ is an analytic function in the
``off-shell" energies $\omega=2v \cdot k$ and $\omega'=2v' \cdot k'$
with discontinuities for positive values of these variables. It also
depends on the velocity transfer $y=v \cdot v'$, which is fixed in a
physical region. $\mathcal {L}_{V,A}$ is the Lorentz structure.
Following the standard process of the QCD sum rule method and
confining us to the operators of dimension $D \leq 5$ in OPE, the
resulting equation reads
\begin{eqnarray}\label{rule2}
\tau(y)f_{-,1/2}f_{+,5/2}e^{-(\bar{\Lambda}_{-,1/2}+\bar{\Lambda}_{+,5/2})/T}&=&
\int^{\omega_{c0}}_{0}\int^{\omega_{c1}}_{0}d\nu d\nu^{'}
e^{-\frac{\nu+\nu^{'}}{2T}} \rho_{pert}(\nu,\nu^{'},y)
\nonumber\\&-&\frac{T^{2}}{3\times2^{4}}\frac{4y+7}{(y+1)^{3}}\langle\frac{\alpha_{s}}{\pi}GG\rangle.
\end{eqnarray}
We have performed a double Borel transformation in $\omega$ and
$\omega^{'}$ on both sides of the sum rule and take the Borel
parameters to be equal \cite{Neu,Hua1}: $T_{1}=T_{2}=2T$. The
perturbative spectral density is
\begin{eqnarray}\label{perturb1}
&\rho_{pert}(\nu,\nu^{'},y)=\frac{3}{2^{9}\pi^{2}}\frac{1}{(y+1)^{\frac{5}{2}}
(y-1)^{\frac{7}{2}}}\nu^{'}[(5\nu-12y\nu^{'}-3\nu^{'})\nu^{2}+(3\nu-\nu^{'})(2y^{2}+2y+1)\nu^{'2}]
\nonumber\\&\times\Theta(\nu)\Theta(\nu^{'})\Theta(2y\nu\nu^{'}-\nu^{2}-\nu^{'2}).
\end{eqnarray}
Following the discussions in Refs. \cite{Neu,Blo}, we must make a
change for the integral variables: $\nu_{-}=\nu-\nu^{'}$,
$\nu_{+}=\frac{\nu+\nu^{'}}{2}$ and choose the triangular region
defined by the bounds: $0\leq \nu_{+}\leq \omega_{c}$,
$-2\sqrt{\frac{y-1}{y+1}}\nu_{+}\leq \nu_{-}\leq
2\sqrt{\frac{y-1}{y+1}}\nu_{+}$. As discussed in Refs.
\cite{Blo,Neu}, the upper limit $\omega_{c}$ for $\nu_{+}$ in the
region
$\frac{1}{2}[(y+1)-\sqrt{y^{2}-1}]\omega_{c0}\leqslant\omega_{c}\leqslant\frac{1}{2}(\omega_{c0}+\omega_{c1})$
is reasonable. After the integral over the ``off-diagonal" variable
$\nu_{-}$ has been done, the final sum rule for $\tau$ appears to be
\begin{eqnarray}\label{rule3}
\tau(y)f_{-,1/2}f_{+,5/2}e^{-(\bar{\Lambda}_{-,1/2}+\bar{\Lambda}_{+,5/2})/T}&=&
\frac{3}{5\times2^{4}\pi^{2}}\frac{1}{(1+y)^{4}}\int^{\omega_{c}}_{0}d\nu_{+}e^{-\frac{\nu_{+}}{T}}
\nu^{5}_{+}
\nonumber\\&-&\frac{T^{2}}{3\times2^{4}}\frac{4y+7}{(y+1)^{3}}\langle\frac{\alpha_{s}}{\pi}GG\rangle.
\end{eqnarray}

Numerical calculations are straightforward. We first estimate the
mass of the final charmed doublet from the two-point sum rule
(\ref{massrule}). The stability window exists when $\omega_{c1}$
lies in the interval $3.5$ to $3.7$ GeV. When $\tau(y)$ is
estimated, the systematic uncertainties can be reduced through
dividing the three-point sum rule (\ref{rule3}) by the square roots
of two-point sum rules (\ref{consrule1}) and (\ref{consrule2}), as
many authors did \cite{Neu,Hua1}. Imposing the usual criteria for
the upper and lower bounds of the Borel parameter, we found that
they have a common sum rule ``window":
$0.7\mbox{GeV}<T<1.5\mbox{GeV}$, which overlaps with that of the
two-point sum rule (\ref{consrule1})[see Fig. 1(a)]. Notice that the
Borel parameter in the sum rule for the three-point correlator is
twice the Borel parameter in the sum rule for the two-point
correlator. In the evaluation we have taken
$2.0\mbox{GeV}<\omega_{c0}<2.4\mbox{GeV}$, which is fixed by
analyzing the corresponding two-point sum rule \cite{Hua1,Neu}.
According to the discussion above, we can fix $\omega_{c}$ in the
region $2.8\mbox{GeV}<\omega_{c}<3.0\mbox{GeV}$. For the QCD
parameters entering the theoretical expressions, we take the
standard values:
$\label{qcond}\langle\overline{q}q\rangle=-(0.24)^{3}
\mbox{GeV}^{3}$, $\label{gcond}\langle\alpha_{s}GG\rangle=0.04
\mbox{GeV}^{4}$, and $\label{mcond}m^{2}_{0}=0.8 \mbox{GeV}^{2}$.
Taking all these into account, the bounding energy of the final
meson doublet is found to be
\begin{equation}\label{mass}
\bar{\Lambda}_{+,5/2}=1.58\pm0.08 \mbox{GeV}.
\end{equation}
The evaluation of $\tau(y)$ is shown in Fig. 1(b). The resulting
curve can be parametrized by the linear approximation
\begin{equation}\label{linear}
\tau(y)=\tau(1)[1-\rho^{2}_{\tau}(y-1)],\text{ }
\tau(1)=0.1\pm0.02,\text{ }\rho^{2}_{\tau}=0.15\pm0.02.
\end{equation}
The errors mainly come from the uncertainty due to $\omega_{c}$ and
$T$. It is difficult to estimate these systematic errors which are
brought in by the quark-hadron duality.
\begin{figure}\begin{center}
\begin{tabular}{ccc}
\begin{minipage}{7cm} \epsfxsize=7cm
\centerline{\epsffile{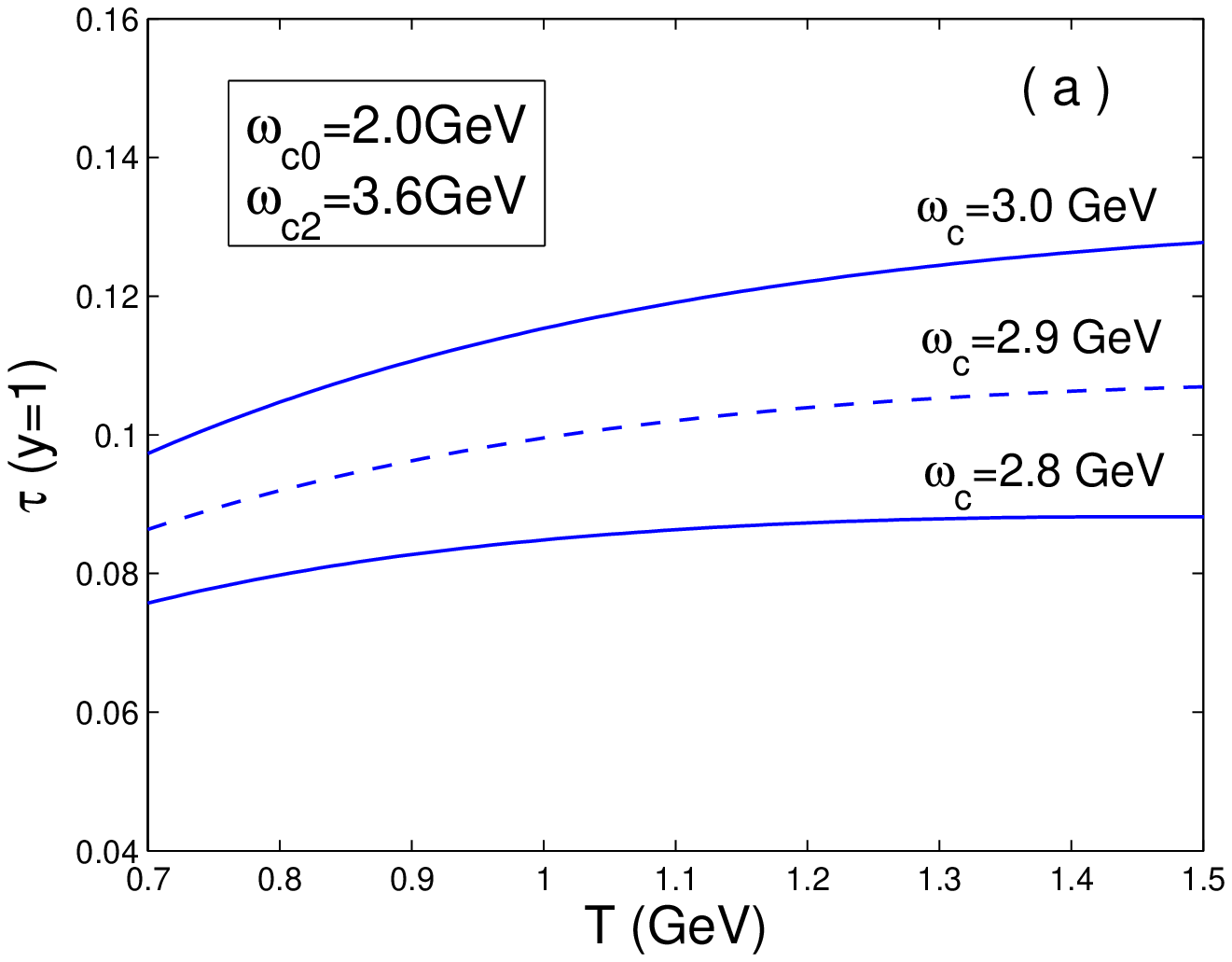}}
\end{minipage}& &
\begin{minipage}{7cm} \epsfxsize=7cm
\centerline{\epsffile{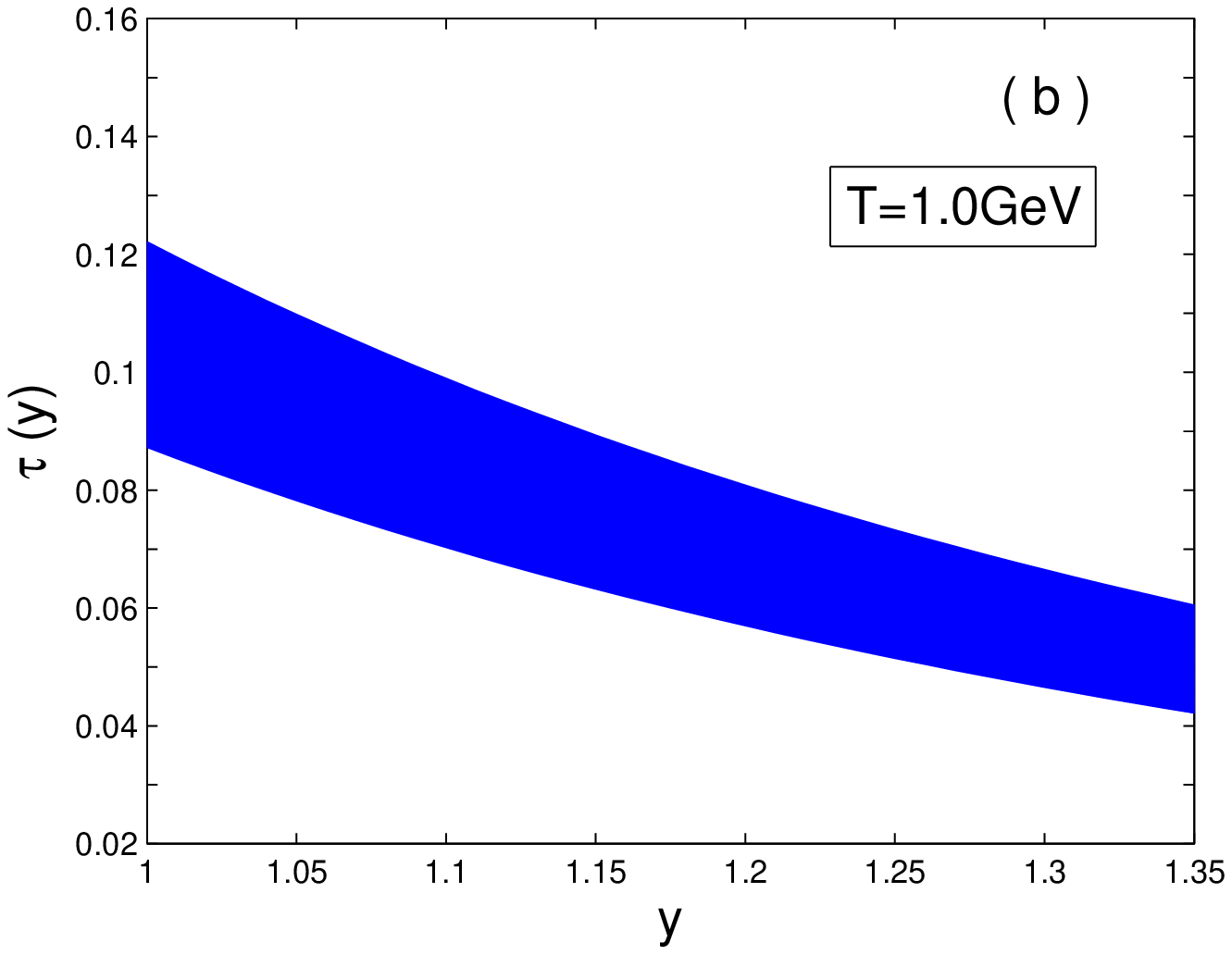}}
\end{minipage}
\end{tabular}
\caption{(a) \it{Dependence of $\tau(y)$ on the Borel parameter $T$
at $y=1$}.(b) \it{Prediction for the Isgur-Wise function $\tau(y)$}.
}
\end{center}
\end{figure}

The differential decay rates are calculated by making use of
formulas (\ref{matrix1}) and (\ref{matrix2}) given above:
\begin{equation}\label{rate3}
\frac{d\Gamma}{dy}(B\rightarrow D_{2}^{*'}\ell\overline{\nu})=
\frac{G^{2}_{F}V^{2}_{cb}m^{2}_{B}m^{3}_{D_{2}^{*'}}}{360\pi^{3}}(\tau(y))^{2}(y+1)^{\frac{5}{2}}
(y-1)^{\frac{7}{2}}[(1+r_{1}^{2})(7y+3)-2r_{1}(4y^{2}+3y+3)],
\end{equation}
\begin{equation}\label{rate4}
\frac{d\Gamma}{dy}(B\rightarrow D_{3}\ell\overline{\nu})=
\frac{G^{2}_{F}V^{2}_{cb}m^{2}_{B}m^{3}_{D_{3}}}{360\pi^{3}}(\tau(y))^{2}(y+1)^{\frac{5}{2}}
(y-1)^{\frac{7}{2}}[(1+r_{2}^{2})(11y-3)-2r_{2}(8y^{2}-3y+3)],
\end{equation}
with $r_{i}=\frac{m_{D^{**}_{i}}}{m_{B}}$ ($D^{**}_{i}= D_{2}^{*'},
D_{3}$ for $i=1, 2$ ). Taking into account that the c-quark mass is
$m_{c}=1.27 \mbox{GeV}$, the mass of the $s^{P}_{l}=\frac{5}{2}^{+}$
charmed doublet appears to be $2.85 \mbox{GeV}$. For the mass of the
initial $B$ meson, we take $5.279\mbox{GeV}$\cite{Pdg}. The maximal
values of $y$ are then $y^{D^{*'}_{2}}_{max} = y^{D_{3}}_{max} =
(1+r_{1,2}^{2})/2r_{1,2}\approx 1.196$. By using the parameters
$V_{cb}=0.04$, $G_{F}=1.166\times10^{-5}\mbox{GeV}^{-2}$, the
semileptonic decay rates of $B\rightarrow (D^{*'}_{2},
D_{3})\ell\overline{\nu}$ turn out to be $1.85\times10^{-19}$GeV and
$1.78\times10^{-19}$GeV, respectively. Considering that
$\tau_{B}=1.638 \text{ps}$ \cite{Pdg}, the branching ratios are
$4.6\times10^{-7}$ and $4.4\times10^{-7}$, respectively. So we can
see that the contributions of the decay modes $B\rightarrow
(D^{*'}_{2}, D_{3})\ell\overline{\nu}$ to the inclusive semileptonic
$B$ decay rate are negligibly small. The severe suppression is
mainly due to the orthogonality of the F-wave orbitally excited
spatial wave function of $D^{**}_{i}$'s to B's. As expected, the
dynamical suppression in these processes is much severer than the
CKM suppression in the $b\rightarrow u\ell\nu$ processes.

In summary, we have estimated the leading-order universal Isgur-Wise
function describing the $B$ meson of ground-state transition into
orbitally excited $F$-wave charmed resonances, the ($2^{+}$,
$3^{+}$) states ($D^{*'}_{2}$, $D_{3}$), which belong to the
$s_{l}^{P}=\frac{5}{2}^{+}$ heavy quark doublet, by use of the QCD
sum rule within the framework of HQET. The semileptonic decay widths
we predict are $1.85\times10^{-19}$GeV and $1.78\times10^{-19}$GeV.
The corresponding branching ratios are $4.6\times10^{-7}$ and
$4.4\times10^{-7}$, respectively. They are much smaller than those
of the $B$ to low-lying charmed meson modes and even the CKM
suppressed modes. Therefore, their contributions to the inclusive
semileptonic $B$ decay rate are nearly negligible. We cannot expect
the high-excited charmed semileptonic modes to contribute much to
the total $B$ decay in the experiments because of the severe
dynamical suppression.

\acknowledgments This work was supported in part by the National
Natural Science Foundation of China under Contract No. 10675167.

\end{document}